\documentclass{article}

\usepackage{PRIMEarxiv}

\usepackage[utf8]{inputenc} 
\usepackage[T1]{fontenc}    
\usepackage{hyperref}       
\usepackage{url}            
\usepackage{booktabs}       
\usepackage{amsfonts}       
\usepackage{nicefrac}       
\usepackage{microtype}      
\usepackage{lipsum}
\usepackage{fancyhdr}       
\usepackage{graphicx}       
\graphicspath{{media/}}     

\newcommand{\logit}{\mathrm{logit}}
 \usepackage{lscape}
 \usepackage{booktabs}
 \usepackage{graphicx}
 \usepackage{lipsum}
 \usepackage{algorithm}
\usepackage{algpseudocode}
\usepackage{amsmath}
\usepackage{prodint}
\usepackage{colortbl}
\usepackage{threeparttable}
\usepackage{tcolorbox}

\title{A general approach to fitting multistate cure models based on an extended-long-format data structure}

\author{
  Yilin Jiang \\
  Leiden University\\
  Princess Maxima Center \\
  the Netherlands\\
  \texttt{y.jiang@math.leidenuniv.nl}\\
\And
 Harm van Tinteren \\
  Princess Maxima Center\\
  the Netherlands\\
  \And
  Marta Fiocco\\
  Leiden University\\
  Leiden University Medical Center\\
  Princess Maxima Center \\
  the Netherlands\\
}

\begin{document}
\maketitle

\begin{abstract}
A multistate cure model is a statistical framework used to analyze and represent the transitions individuals undergo between different states over time, accounting for the possibility of being cured by initial treatment. This model is particularly useful in pediatric oncology where a proportion of the patient population achieves cure through treatment and therefore will never experience certain events. Despite its importance, no universal consensus exists on the structure of multistate cure models. Our study provides a novel framework for defining such models through a set of non-cure states. We
develops a generalized algorithm based on the extended long data format, an extension of the traditional long data format, where
a transition can be divided into two rows, each with a weight assigned reflecting the posterior probability of
its cure status. The multistate cure model is built upon the current framework of multistate model and mixture cure model. The proposed algorithm makes use of the Expectation-Maximization (EM) algorithm and weighted likelihood representation such that it is easy to implement with standard packages. Additionally, it facilitates dynamic prediction. The algorithm is applied on data from the European Society for Blood and Marrow Transplantation (EBMT). Standard errors of the estimated parameters in the EM algorithm are obtained via a non-parametric bootstrap procedure, while the method involving the calculation of the second-derivative matrix of the observed log-likelihood is also presented.
\end{abstract}


\section{Introduction}\label{sec1}

As medical technology advances, there has been a growing interest in integrating the concept of the cured proportion into multistate models.
This is particularly useful in pediatric oncology where a proportion of the patients achieves cure through treatment and therefore one or more of disease states may never occur for them. It allows researchers to quantify the likelihood of transitions between states, examine the factors influencing these transitions while considering the probability of being cured and estimating the curative effect of a treatment. It can therefore provide a more comprehensive understanding of the disease progression or remission patterns being studied and aid in personalized treatment decisions and prognosis prediction.

Over the past few years, various frameworks have been brought forward to incorporate cured fraction into multistate models. Sommer et al.\cite{sommer2017time} proposed a multistate cure-death model where cure and death were studied simultaneously as coprimary endpoints. In their model, clinical cure was recorded through a test-of-cure visit. However, in most situations cure status is a latent variable that cannot be measured. The classic mover-stayer model, which mainly deals with longitudinal data, approaches the unobserved population heterogeneity through a two-class mixture model. It was first introduced by Blumen, Kogan and McCarthy \cite{blumen1955industrial} and Goodman \cite{goodman1961statistical}. They assumed the population consisted of two groups: stayers remaining in certain states and movers following ordinary Markov processes. Mover-stayer models focus on estimating the proportion of stayers and the transition probabilities for movers. Sundin et al. \cite{sundin2023semi} borrowed the idea from the mover-stayer model and constructed a semi-Markov multistate cure model in which cured proportions accounted for those who will never transition out of a certain state. Conlon et al.\cite{conlon2014multi} and Beesley and Taylor\cite{beesley2019algorithms} developed multistate cure models based on the illness-death model where the "healthy" state was divided into two baseline states, cured and non-cured.  

In terms of the estimation method, some employed the Bayesian Monte Carlo Markov chain (MCMC) techniques \cite{sundin2023semi,conlon2014multi}. Given the intricate nature of multistate cure models, a comprehensive specification of statistical priors is necessary. Beesley presented an Expectation-Maximization (EM) algorithm. With careful thought, they constructed an augmented data structure wherein each subject had rows for all model transitions
\cite{beesley2019algorithms}.

In our proposal, the multistate cure model is defined via a set of non-cure states, e.g. recurrence of a disease and metastasis. Subjects observed to undergo any transition to or from these non-cure states are certain to be non-cured at baseline. Here cured or non-cured is a baseline status, rather than an entry state. This approach is more intuitive and logical as cure status is often latent or unobservable and in practice a subject should not have transition path with uncertainty.

Our proposed algorithm utilizes an EM algorithm and weighted likelihood representation such that it is easy to implement with standard software and existing packages. Previous literature primarily focused on the illness-death model, thereby limiting its applicability. We introduce a particularly useful way of representing the data, which we call extended long data format, as an extension of the commonly used long data format in multistate models, which is straightforward and renders our proposed algorithm capable of fitting any multistate cure model.

The algorithm is applied to data from the European Society for Blood and Marrow Transplantation (EBMT) registry. The article is organized as follows. In Section \ref{sec2}, we provide an overview of the EBMT data and describe the multistate cure model notation and specification. In Section \ref{sec3}, we outline the likelihood construction and discuss the EM algorithm in this setting. The extended long data format and pseudocode illustrating our proposed algorithm are presented in Section \ref{sec4}. We then discuss how to obtain the standard errors of the parameters involved in the EM algorithm in Section \ref{sec5}. An extension to dynamic prediction is demonstrated in Section \ref{sec_new}. We apply our proposed method on the EBMT data and present the results in Section \ref{sec6}. The article ends with a discussion in Section \ref{sec7}.

\section{Multistate cure model notation and specification}\label{sec2}
\subsection{EBMT data}

For the purpose of illustration, here we consider survival of leukemia patients after bone marrow transplantation. The data, originated from the EBMT, is available in the R package \textbf{mstate} as \emph{ebmt4} in wide format, with each row corresponding to an individual patient. The data frame consists of 2279 acute lymphoid leukemia patients who had an allogeneic bone marrow transplant at the EBMT between 1985 and 1988. Detailed description about the data can also be found in previous studies by Fiocco, Putter and van Houwelingen \cite{fiocco2008reduced}, van Houwelingen and Putter \cite{van2008dynamic,van2011dynamic} and de Wreede, Fiocco and Putter \cite{de2011mstate}. Please note that, as explained by de Wreede et al\cite{de2011mstate}, small adjustments have been made to make the data suitable for a multistate analysis and intermediate events have been abstracted from the actual disease to avoid misinterpretation. Additionally, twenty-three relapsed patients were lost to follow-up for unclear reasons.  Therefore, clinical interpretation based on the analysis result should be treated with caution.

\begin{figure*}[h]
\centerline{\includegraphics[width=\textwidth,height=22pc]{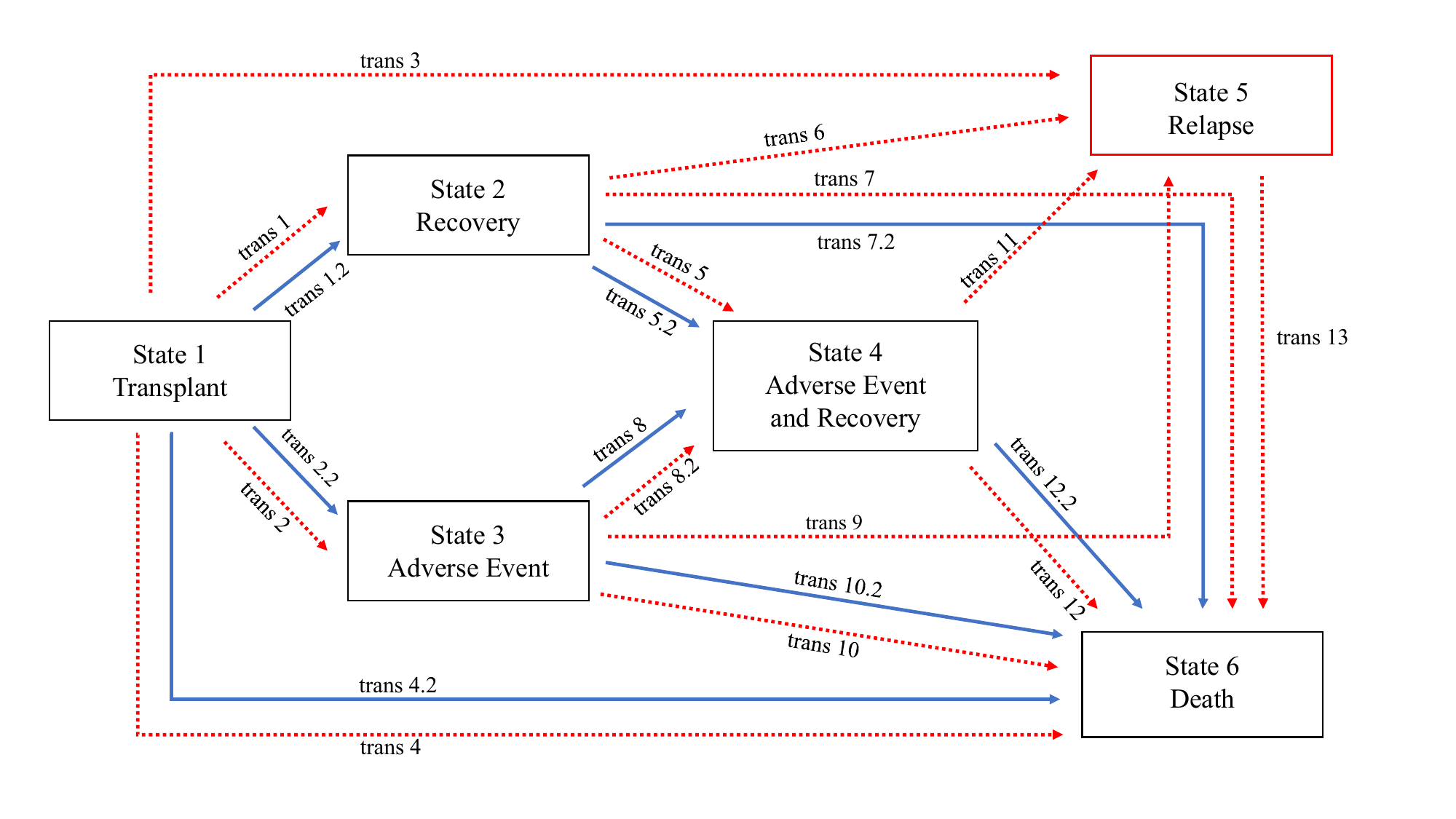}}
\caption{Diagram of the six-state model for leukemia patients after bone marrow transplant (ebmt4).
The red square represents the non-cure state: relapse. Red dashed lines indicate transitions corresponding to non-cured status, whereas blue solid lines refer to transitions corresponding to cured status.\label{diagram}}
\end{figure*}

Figure \ref{diagram} presents a diagram depicting the EBMT data via a six-state model. All subjects were transplanted in first complete remission, entering the initial state 1, the condition of being alive and in remission with no recovery or adverse event. Subjects enter state 2 upon recovery and enter state 3 upon adverse event. Here, recovery is defined as the recovery of platelet counts to a level of $>20\times10^9/L$ and an adverse event refers to acute graft-versus-host disease of grade 2 or higher. State 4 represents the condition of being alive with both recovery and adverse event having occurred. Relapse (state 5)  indicates treatment failure. State 6, the absorbing state, represents death. 

Our proposed multistate cure model is defined through a single non-cure state, namely the relapse state, indicated by red square in figure \ref{diagram}. Here, cure is defined as no relapse after transplant.
To address the population heterogeneity in terms of cured and non-cured baseline status, we use blue solid lines for transitions followed by cured patients and red dashed lines for transitions followed by non-cured patients.

Table \ref{baseline char} shows the patient characteristics. Four prognostic factors at baseline are considered as covariates: year of transplant (factor with 3 levels: 1985-1989, 1990-1994,1995-1998), age at transplant in years (factor with 3 levels: $\leq$ 20, 20-40, >40), prophylaxis (factor with 2 levels: no and yes) and donor-recipient gender match (factor with 2 levels: no gender mismatch and gender mismatch). Among the 2279 leukemia patients, 370 experienced relapses, with 305 of these patients subsequently passing away. In total, there were 838 deaths following bone marrow transplant.

\subsection{Notation}

We consider a continuous time Markov model with finite state space $\mathbb{S} = \{1,...,s\}$. Let $V(t)$ denote the state occupied at time $t$ and $\{V(t), t\geq0\}$ the stochastic process. Here $t$ refers to the time since start of study, i.e., clock-forward approach is chosen for the time scale. With a Markov process assumed, the dependence of $V(t)$ on the history is only through the currently occupied state.

Under the multistate cure model setting, we need to define a set of non-cure states, $\mathbb{S}_{nc} \subseteq \mathbb{S}$. Let $G$ be the latent variable for baseline cure status, $G=0$ meaning non-cured at baseline and $G=1$ cured at baseline. Only subjects who do not enter any of the non-cure states have the possibility of being cured at baseline. Subjects who undergo transitions to or from the non-cure states are known to have non-cured status at baseline. The probability of being cured at baseline is modelled through logistic regression. Denote by $Z_{\alpha}$ and $\alpha$ the vector of covariates and corresponding regression coefficients for the cure probability model respectively, we have
\begin{equation}  \label{logit}
\logit\bigl(P(G=1\    | \, Z_{\alpha})\bigr)=Z_{\alpha}^T \alpha, \qquad   \textrm{where} \  P(G=0\   |\, Z_{\alpha})+P(G=1\   |\, Z_{\alpha})=1.
\end{equation}

Let $N_{kl}(t)$ denote a right-continuous function that counts the number of transitions from state $k$ to $l$ over the time interval $[0,t]$. Then $\Delta N_{kl}(t)=N_{kl}(t+\Delta t-)-N_{kl}(t-)$ represents the number of $k \rightarrow l$ transitions over $[t,t+\Delta t)$ and $dN_{kl}(t)=lim_{\Delta t \rightarrow 0}\Delta N_{kl}(t)=1$ indicates a $k \rightarrow l$ transition occurs at time t, where time $t-$ means just before time $t$.
Denote by $Y_{k}(t)=\mathrm{I}(V(t)=k)$ the indicator that a subject is in state $k$ at time $t$, i.e., the at-risk process. The intensity function in counting process notation is as follows 
\begin{equation}    \label{intensity1}
{P\bigl(dN_{kl}(t)=1\bigr)} \approx Y_{k}(t-)\lambda_{kl}(t)dt, \qquad \textrm{for all} \  k \neq l ,
\end{equation} 
where the hazard rate for $k \rightarrow l$ transition at time $t$ is defined as
\begin{equation}  \label{intensity2}
\lambda_{kl}(t)=\lim_{\Delta t \rightarrow 0} \frac {P\bigl(V(t+\Delta t-)=l\    |\, V(t-)=k\bigr)} {\Delta t}.
\end{equation} 

Assume proportional hazards (PH) for each $k \rightarrow l$ transition. Through Cox proportional hazards model, we estimate the hazard rate via
\begin{equation}\label{cox}
\lambda_{kl}(t\ |\, Z_{\beta}, G ) = \lambda_{0,kl}(t)\ \exp(Z_{\beta}^T \beta_{kl})\ \exp(G\gamma_{kl}),
\end{equation}
where $Z_{\beta}$ and $\beta_{kl}$ denote the vector of covariates and the $k \rightarrow l$ transition-specific regression coefficients for the survival analysis.  The term $\lambda_{0,kl}$ is the baseline hazard for $k \rightarrow l$ transition. The baseline cure status $G$ is incorporated in a multiplicative fashion. For transitions in which either $k$ or $l$ is the non-cure state $s_{nc}$, baseline cure status $G=0$ and the last term $\exp(G\gamma_{kl})=1$. 

Let $\mathcal{Z}=\{Z_{\alpha},Z_{\beta}\}$ denote the covariates for the analysis. In the EBMT example, they are all fixed at baseline. Let $\boldsymbol{\theta} =\{\alpha, \boldsymbol{\beta}, \boldsymbol{\gamma}, \boldsymbol{\lambda_0(\cdot)} \}$ represent the model parameters where $\alpha$ is for the cure probability analysis, $\boldsymbol{\beta}$, $\boldsymbol{\gamma}$ and the baseline hazard function $\boldsymbol{\lambda_0(\cdot)}$ for the survival analysis. Note that $\boldsymbol{\beta}$, $\boldsymbol{\gamma}$ and $\boldsymbol{\lambda_0(\cdot)}$ are a collection of transition-specific parameters. Furthermore, we assume independent censoring for simplicity, i.e., censoring times $\mathcal{C}$ are independent of all underlying transition times given covariates.

\section{Likelihood construction and EM algorithm}\label{sec3}

\subsection{Likelihood construction}
For subject $i$, $i=1,...,n$, we assume to observe  $\mathbb{D}_{i}=\{(N_{i,kl}(t),Y_{i,k}(t)): k,l \in \mathbb{S}$ and $k \neq l$\} over the time interval  $[0, X_i]$ where $X_i$ is the censoring time $C_i$ or the time point $T_i$ when an absorbing state is reached, whichever comes first, i.e., $X_i=min(C_i, T_i)$. With the baseline cure status $G_i$ and the covariates $Z_{i,\beta}$, the probability for subject $i$ experiencing the observed path is
\begin{equation}\label{likelihood per subject given G}
P(\mathbb{D}_{i} \,|\,Z_{i,\beta},G_i) =\prod_{k=1}^{s}\prod_{l \neq k=1}^{s} \Bigl\{ \Prodi_{t=0}^{X_i}
\bigl(\lambda_{i,kl}(t\,|\,Z_{i,\beta},G_i)dt\bigr)^{dN_{i,kl}(t)}\Bigr\}\,\mathrm{exp}\Bigl(-\int_{0}^{X_i}Y_{i,k}(u^-)\lambda_{i,kl}(u\,|\,Z_{i,\beta},G_i)du \Bigr).
\end{equation}

The \emph{observed data likelihood} for subject $i$ is as follows
\begin{equation}\label{ob likelihood per subject}
P(\mathbb{D}_{i}\,|\,\mathcal{Z}_{i})=\sum_{g=0}^{1} P(\mathbb{D}_{i} \,|\,Z_{i,\beta},G_i=g)P(G_i=g |\,Z_{i,\alpha}),
\end{equation}

\noindent where $G_i$ is unobserved and $Z_{i,\alpha}$ are  the baseline covariates.
The overall observed data log-likelihood takes the form of $\ell_{O}(\boldsymbol{\theta})=\sum_{i=1}^{n}\mathrm{log}P(\mathbb{D}_{i}\,|\,\mathcal{Z}_{i})$. This now becomes a maximum-likelihood estimation (MLE) problem. As it is hard to get expressions for the derivatives needed by gradient methods, we consider the expectation-maximization (EM) algorithm.

\subsection{EM algorithm}\label{em basics}
EM algorithm is a powerful approach to optimize the overall observed data log-likelihood $\ell_{O}(\boldsymbol{\theta})$. It makes use of the \emph{complete data log-likelihood} $\ell_{C}(\boldsymbol{\theta})$, defined as if $G$ was observed,
\begin{equation}\label{com likelihood}
\ell_{C}(\theta)=\sum_{i=1}^{n} \mathrm{log}P(\mathbb{D}_{i},G_i\,|\,\mathcal{Z}_{i})=\sum_{i=1}^{n}\sum_{g=0}^{1}\mathrm{I}(G_i=g)\bigl[\mathrm{log}P(\mathbb{D}_{i} \,|\,Z_{i,\beta},G_i=g)+\mathrm{log} P(G_i=g |\,Z_{i,\alpha})\bigr].
\end{equation}

The EM algorithm is as follows:
\begin{tcolorbox}
{\noindent\textbf{Initialize} Start with an initial estimate of the model parameters $\boldsymbol{\theta} =\{\alpha,\boldsymbol{\beta},\boldsymbol{\gamma}, \boldsymbol{\lambda_0(\cdot)} \}$. \\
Until the increase of the observed data log-likelihood is sufficiently small, i.e., $\ell_O(\boldsymbol{\theta}^{(r+1)})-\ell_O(\boldsymbol{\theta}^{(r)}) < \epsilon$,\\
\textbf{E-step} Given the most recent parameter estimate $\boldsymbol{\theta}^{(r)}$, calculate the expected complete data log-likelihood, 
\begin{equation}\label{expected com likelihood}
Q(\boldsymbol{\theta}|\boldsymbol{\theta}^{(r)})=E_{\boldsymbol{\theta}^{(r)}}[\ell_{C}(\boldsymbol{\theta})\,|\, \mathbb{D}_{i}].
\end{equation}
\textbf{M-step} Find $\theta=\underset{\boldsymbol{\theta}}{\arg\max} Q(\boldsymbol{\theta}|\boldsymbol{\theta}^{(r)})$.}
\end{tcolorbox}
We can rewrite \eqref{expected com likelihood} as
$
Q(\boldsymbol{\theta}|\boldsymbol{\theta}^{(r)})=Q_{1}(\boldsymbol{\beta},\boldsymbol{\gamma},\boldsymbol{\lambda_0(\cdot)}\,|\,\boldsymbol{\theta}^{(r)})+Q_{2}(\alpha\,|\,\boldsymbol{\theta}^{(r)})
$, where 
\begin{equation}\label{Q1}
Q_{1}(\boldsymbol{\beta},\boldsymbol{\gamma},\boldsymbol{\lambda_0(\cdot)}\,|\,\boldsymbol{\theta}^{(r)})=\sum_{i=1}^{n}\Bigl\{ E_{\boldsymbol{\theta}^{(r)}}(G_{i}\,|\, \mathbb{D}_{i} )\,\mathrm{log}P_{\boldsymbol{\theta}}(\mathbb{D}_{i} \,|\, G_{i}=1, Z_{i,\beta})+\bigl(1-E_{\boldsymbol{\theta}^{(r)}}(G_{i}\,|\, \mathbb{D}_{i})\bigr)\,\mathrm{log}P_{\boldsymbol{\theta}}(\mathbb{D}_{i}\, |\, G_{i}=0, Z_{i,\beta})\Bigr\},
\end{equation}
and
\begin{equation}\label{Q2}
Q_{2}(\alpha\,|\,\boldsymbol{\theta}^{(r)})=\sum_{i=1}^{n}\Bigl\{ E_{\boldsymbol{\theta}^{(r)}}(G_{i}\,|\, \mathbb{D}_{i})\,\mathrm{log}P_{\boldsymbol{\theta}}(G_{i}=1 \,|\,  Z_{i,\alpha})+\bigl(1-E_{\boldsymbol{\theta}^{(r)}}(G_{i}\,|\, \mathbb{D}_{i})\bigr)\,\mathrm{log}P_{\boldsymbol{\theta}}(G_{i}=0\, |\, Z_{i,\alpha})\Bigr\}.
\end{equation}

Let $w_{i}^{(r)}$ denote $E_{\boldsymbol{\theta}^{(r)}}(G_{i}\,|\, \mathbb{D}_{i})$ and $\pi_{i}^{(r)}$ denote $P_{\boldsymbol{\theta}^{(r)}}(G_{i}=1\,|\,Z_{i,\alpha})$ for simplicity. We can obtain $w_{i}^{(r)}$ using Bayes' rule,
\begin{equation}\label{w}
\begin{split}
w_{i}^{(r)}& =P_{\boldsymbol{\theta}^{(r)}}(G_{i}=1 \,|\,\mathbb{D}_{i})\\
& =\frac{\pi_{i}^{(r)}P_{\boldsymbol{\theta}^{(r)}}\bigl(\mathbb{D}_{i} \,|\, G_{i}=1, Z_{i,\beta}\bigr)}{\pi_{i}^{(r)}P_{\boldsymbol{\theta}^{(r)}}\bigl(\mathbb{D}_{i} \,|\, G_{i}=1, Z_{i,\beta}\bigr)+(1-\pi_{i}^{(r)})P_{\boldsymbol{\theta}^{(r)}}\bigl(\mathbb{D}_{i} \,|\, G_{i}=0, Z_{i,\beta}\bigr)}.
\end{split}
\end{equation}

The separation of $\boldsymbol{\beta}$, $\boldsymbol{\gamma}$, $\boldsymbol{\lambda_0}$ and $\alpha$ in $Q(\boldsymbol{\theta}|\boldsymbol{\theta}^{(r)})$ makes the M-step relatively straightforward. Worthy to note that the structure of $Q_{1}(\boldsymbol{\beta}, \boldsymbol{\gamma}, \boldsymbol{\lambda_0(\cdot)}\,|\,\boldsymbol{\theta}^{(r)})$ resembles the log-likelihood for a weighted multistate model, while the form of $Q_{2}(\alpha\,|\,\boldsymbol{\theta}^{(r)})$ is akin to the log-likelihood for a weighted logistic regression. The weight $w_{i}^{(r)}$, reflecting the posterior probability of the cure status for subject $i$, gets updated in every E-step.

\section{Algorithm Implementation}\label{sec4}
\subsection{Data preparation}
To fit our proposed EM algorithm, data in its original wide format needs to be recoded. Table \ref{data wide} shows an example from the EBMT data, with one row per subject. Subject 1 experienced an adverse event at 60 days after transplant and died at 143 days. Subject 2 experienced an adverse event at 12 days after transplant, recovered at 29 days, had a relapse at 422 days and died at 579 days. Subject 3 had an adverse event at 14 days after transplant and got censored at 3687 days.

Since $\boldsymbol{\beta}$, $\boldsymbol{\gamma}$, $\boldsymbol{\lambda_0}$ and $\alpha$ can be estimated separately in the M-step, data preparation will be illustrated from two perspectives. 

\subsubsection{$Q_{1}(\boldsymbol{\beta}, \boldsymbol{\gamma}, \boldsymbol{\lambda_0(\cdot)}\,|\,\boldsymbol{\theta^{(r)}})$-focused data format}\label{q1data}
The form of $Q_{1}(\boldsymbol{\beta}, \boldsymbol{\gamma}, \boldsymbol{\lambda_0(\cdot)}\,|\,\boldsymbol{\theta^{(r)}})$ (see formula \eqref{Q1}) resembles the log-likelihood for a weighted multistate model. With this in mind, we introduce an extended long data format, an extension of the long data format commonly used in the field of multistate models\cite{putter2007tutorial}.

Table \ref{data long-long} shows the extended long data format for the subjects in table \ref{data wide}. Similar to long data format, each row represents a subject at risk for a certain transition indicated by column \texttt{trans}, with a \texttt{from} and a \texttt{to} column specifying from which state the transition starts and to which state it ends. Columns \texttt{Tstart} and \texttt{Tstop} specify the starting and stopping time for the subject being at risk for a specific transition and column \texttt{status} shows whether or not the \texttt{to} state was reached. Different from conventional long data format, the same \texttt{from$\rightarrow$to} transition can be divided into two rows, depending on whether the non-cure state is concerned. As shown in figure \ref{diagram}, there are two types of lines for transitions: $1\rightarrow2$,  $1\rightarrow3$, $1\rightarrow6$, $2\rightarrow4$, $2\rightarrow6$, $3\rightarrow4$, $3\rightarrow6$ and $4\rightarrow6$. The red dashed lines refer to transitions corresponding to non-cured status, while the blue solid lines indicate transitions corresponding to cured status. The distinction is also reflected in column \texttt{trans} where transitions corresponding to non-cured status are labelled as an integer while for transitions corresponding to cured status we add ".2".  Extra columns named with \texttt{cure} are added for these \texttt{from$\rightarrow$to} pairs that may have two possibilities of baseline cure status. Columns \texttt{cure12}, \texttt{cure13}, 
\texttt{cure16}, 
\texttt{cure24}, \texttt{cure26}, 
\texttt{cure34}, \texttt{cure36} and \texttt{cure46} represent variable $G$ in \eqref{cox} per transition. The values for time-fixed covariates (year of transplant, age at transplant, prophylaxis and donor-recipient gender match),  are replicated for each patient.

Column \texttt{cumHaz} contains the estimated cumulative hazards from \texttt{Tstart} to \texttt{Tstop} for a specific transition \texttt{trans} given the baseline covariates and cure status. It corresponds to the term $\mathrm{exp}\Bigl(-\int_{0}^{X_i}Y_{i,k}(u^-)\lambda_{i,kl}(u\,|\,Z_{i,\beta},G_i)du\Bigr)$ in \eqref{likelihood per subject given G} for subject $i$ at $k\rightarrow l$ transition, which is 0 outside the interval (\texttt{Tstart}, \texttt{Tstop}] of the relevant row. Column \texttt{hazard} contains the estimated $k\rightarrow l$ transition hazard at \texttt{Tstop}, which is the same as $\lambda_{i,kl}(t\,|\,Z_{i,\beta},G_i)dt$ in \eqref{likelihood per subject given G} where $t=\texttt{Tstop} $.   The probability of subject $i$ being at risk for $k\rightarrow l$ transition and experiencing (or not experiencing, depending on column \texttt{status}) the event is calculated in the column \texttt{likelihood}. This can be interpreted as the likelihood contribution per row, given the baseline cure status. Column \texttt{pi}, representing $\pi$, contains the probability that a subject was cured at baseline. Column \texttt{weight}, corresponding to the weight terms before the log-likelihood in \eqref{Q1}, can be obtained through \eqref{w} utilizing columns \texttt{likelihood}, \texttt{pi}, and \texttt{cure}. The value is $P_{\boldsymbol{\theta}^{(r)}}(G_{i}=1 \,|\,\mathbb{D}_{i})$ if any \texttt{cure} column includes a 1, otherwise, it is $P_{\boldsymbol{\theta}^{(r)}}(G_{i}=0 \,|\,\mathbb{D}_{i})$. For subjects with unknown baseline cure status, columns \texttt{cumHaz}, \texttt{hazard}, \texttt{likelihood}, \texttt{pi} and \texttt{weight} get updated with each iteration of the EM algorithm. For subjects going through the non-cure state (e.g., subject 2), data are expanded as in the long format, since the baseline cure status is known and $w_{i}^{(r)}$ is always 0. As shown in table \ref{data long-long}, for subject 2, only the rows representing non-cured status are kept with column \texttt{weight} being 1. Columns \texttt{cumHaz}, \texttt{hazard}, \texttt{likelihood} and \texttt{pi}, though no updates are involved, are required for the overall observed data log-likelihood calculation at each iteration. When \texttt{status} is 0, any value can be filled in the column \texttt{hazard} since it is not used in the calculation. For computational efficiency, the model fit for transition $5\rightarrow6$ can be performed separately from the EM algorithm, as all the patients reaching state 5 (relapse) have their baseline cure status known.

\subsubsection{$Q_{2}(\alpha|\boldsymbol{\theta}^{(r)})$-focused data format}\label{q2data}

The form of $Q_{2}(\alpha|\boldsymbol{\theta}^{(r)})$ (see formula \eqref{Q2}) can be viewed as the log-likelihood for a weighted logistic regression with $G$ being the binary outcome. Therefore, we have the following data format (table \ref{data q2 focused}) to assist the estimation of parameter $\alpha$. Here column \texttt{cure} is the independent variable. Each subject can have up to two rows representing the different situations of baseline cure status, $G=0$ or 1. Column \texttt{weight}, corresponding to the weight terms before the log-likelihood in \eqref{Q2}, is synchronized with \texttt{weight} in table \ref{data long-long} in every E-step. After model fitting in the M-step, we can then calculate $P_{\alpha^{(r)}}(G_{i}=1\,|\,Z_{i,\alpha})$ with the most recent estimate of $\alpha$, which is essential for obtaining the weight $w_{i}^{(r+1)}$ in the following E-step of the next iteration. Note that the covariates here ($Z_{\alpha}$) can be different from the covariates in table \ref{data long-long} ($Z_{\beta}$).

\subsection{Pseudocode}
To implement the EM algorithm as described in Section \ref{em basics}, we provide the following pseudocode (Algorithm \ref{pseudocode}) to illustrate the iterative procedure. In every E-step, we update the weight $w_i^{(r)}$ for each subject $i$, which consists of calculating the two terms $\pi_i^{(r)}$ and $P_{\boldsymbol{\theta}^{(r)}}\bigl(\mathbb{D}_{i} \,|\, G_{i}, Z_{i,\beta}\bigr)$ as shown in \eqref{w}, based on the previous model fit from the M-step. Every M-step is composed of two sub-steps: fit a weighted Cox model based on Q1-focused data as discussed in Section \ref{q1data} and fit a weighted logistic regression based on Q2-focused data as illustrated in Section \ref{q2data}. The overall observed data log-likelihood is calculated at the end of each iteration. The EM algorithm ends till it stabilizes. For the EBMT data, the convergence threshold $\epsilon$ was equal to 0.0001.
\begin{algorithm}
\caption{Pseudocode for the EM algorithm implementation}\label{pseudocode}
\begin{algorithmic}
\Require Initial parameter estimates $\boldsymbol{\theta}^{(0)}= \{\alpha^{(0)},\boldsymbol{\beta}^{(0)},\boldsymbol{\gamma}^{(0)},\boldsymbol{\lambda_0(\cdot)}^{(0)}\}$ and convergence threshold $\epsilon$

\State Initialize parameter estimates $\boldsymbol{\theta}\leftarrow\boldsymbol{\theta}^{(0)}$, prepare Q1-focused data and Q2-focused data
\Repeat
    \State\textbf{E-step}: 
    \State Update column \texttt{pi} in Q1-focused table using $\alpha^{(r)}$ based on previous weighted logistic model (see \eqref{logit});
    \State Calculate cumulative hazard, instant hazard rate and row likelihood in Q1-focused table using $\boldsymbol{\beta}^{(r)}$, $\boldsymbol{\gamma}^{(r)}$ and $\boldsymbol{\lambda_0(\cdot)}^{(r)}$ based on previous weighted Cox model (see \eqref{cox} and \eqref{likelihood per subject given G});
    \State Calculate $w^{(r)}$ based on columns \texttt{likelihood} and \texttt{pi} in Q1-focused table (see \eqref{w});
    \State Fill in column \texttt{weight} in Q1-focused and Q2-focused tables based on $w^{(r)}$

    \State\textbf{M-step}: 
    \State Fit weighted Cox model using Q1-focused table to obtain $\boldsymbol{\beta}^{(r+1)}$, $\boldsymbol{\gamma}^{(r+1)}$ and $\boldsymbol{\lambda_0(\cdot)}^{(r+1)}$;
    \State Fit weighted logistic model using Q2-focused table to obtain $\alpha^{(r+1)}$

    \State \textbf{Check for convergence}: 
    \State Calculate the observed data log-likelihood for every subject and sum them to obtain $\ell_{O}(\boldsymbol{\theta}^{(r+1)})$ (see \eqref{ob likelihood per subject})   
    \State Update $\boldsymbol{\theta} \leftarrow \boldsymbol{\theta}^{(r+1)}$
\Until{$\ell_O(\boldsymbol{\theta}^{(r+1)})-\ell_O(\boldsymbol{\theta}^{(r)}) < \epsilon$}

\end{algorithmic}
\end{algorithm}

\section{Standard error estimation}\label{sec5}
Due to the uncertainties underlying the weight in the Q1-focused and Q2-focused data, the standard errors derived from the final model fitting of the EM algorithm are underestimated. There are multiple ways to tackle this. Louis\cite{louis1982finding} and Oakes\cite{oakes1999direct} derived explicit formulas for the observed information matrix when using the EM algorithm. Details about the calculation can be found in the Appendix.

However, considering the large number of model parameters included in the multistate cure model for the EBMT data, we chose the non-parametric bootstrap approach to obtain the standard errors. For the convergence threshold, the EM algorithm stops when the increase of the observed data log-likelihood is below 0.01, i.e., $\epsilon = 0.01$, or after 80 iterations. The results are based on 1000 bootstrap samples.

\section{Dynamic prediction}\label{sec_new}
Dynamic prediction is an important tool for individualized prognosis over time. In this section, we show how to estimate the probabilities a subject will be in a certain state at a specific time using the available information within the multistate cure model framework.

Let $H(s)=\{V(u),0\leq u\leq s\}$ denote the history of state occupancy over $[0, s]$. Assuming the current state is $l$, $H(s)$ can be rewritten as $H(s)=\{H(s-),V(s)=l\}$. With the Markov property, we can obtain the probability of being in state $m$ at time $t$, given the event history $H(s)$ and covariates $\mathcal{Z^*}$, via
\begin{equation}\label{predict}
    \begin{split}
        P_{\boldsymbol{\hat{\theta}}}(V(t)=m \,|\, H(s), \mathcal{Z^*})&
        =\sum_{g=0}^{1}P_{\boldsymbol{\hat{\theta}}}(V(t)=m \,|\, H(s-), V(s)=l,\mathcal{Z^*},G=g)P_{\boldsymbol{\hat{\theta}}}(G=g \,|\, H(s), \mathcal{Z^*})\\&
        =\sum_{g=0}^{1}P_{\boldsymbol{\hat{\theta}}}(V(t)=m \,|\,  V(s)=l,\mathcal{Z^*},G=g)P_{\boldsymbol{\hat{\theta}}}(G=g \,|\, H(s), \mathcal{Z^*}),
    \end{split}
\end{equation}
where $\boldsymbol{\hat{\theta}}$ represents the final parameter estimates.
The term $P_{\boldsymbol{\hat{\theta}}}(V(t)=m \,|\,  V(s)=l,\mathcal{Z^*},G=g)$ in \eqref{predict}, also known as the transition probability matrix, can be calculated using the \texttt{probtrans} function from the \textbf{mstate} package, for which one first needs to create an object of class "\texttt{msfit}" containing the estimated cumulative hazards at all event times, $\hat{\lambda}_{kl}(t\,|\,\mathcal{Z_{\beta}^*},G)$ for all $k\rightarrow l$ transitions as in \eqref{cox}. For details about the use of \textbf{mstate} see de Wreede et al. (2011)\cite{de2011mstate}. The term $P_{\boldsymbol{\hat{\theta}}}(G=g \,|\, H(s), \mathcal{Z^*})$ simplifies to $P_{\boldsymbol{\hat{\theta}}}(G=g\,|\,\mathcal{Z_{\alpha}^*})$, as in \eqref{logit}, when only the baseline covariates are known; it is the same as $P_{\boldsymbol{\hat{\theta}}}(G=g \,|\,\mathbb{D})$, as in \eqref{w}, when follow-up information is included. In either case, one can organize the available information into the extended long format and efficiently compute this term with an E-step using the parameter estimates from the very last iteration.

\section{Application to EBMT data}\label{sec6}
 Due to the limited number of events for transition from state 2 to 6, the covariate effect of cure $\gamma_{26}$ was not estimable and therefore was removed from the model. 
 Table \ref{results} presents the results of the multistate cure model fit to the EBMT data. Later year of transplant was associated with higher recovery rates (significant at transition $1\rightarrow2$ and $3\rightarrow4$), lower relapse rates (significant at transition $4\rightarrow5$) and mortality rates (significant at transition $2\rightarrow6$). Older age at transplant was associated with higher mortality rates (significant at transition $1\rightarrow6$ and $4\rightarrow6$). Donor-recipient gender mismatch was associated with higher rates of relapse (significant at transition $2\rightarrow5$)  and death (significant at transition $4\rightarrow6$). None of the baseline covariates included in the model were found to be significant in discriminating between cured patients and non-cured patients. Baseline cure status also did not have a significant impact on disease progression. 

\begin{figure*}[h]
\centerline{\includegraphics[width=0.6\textwidth]{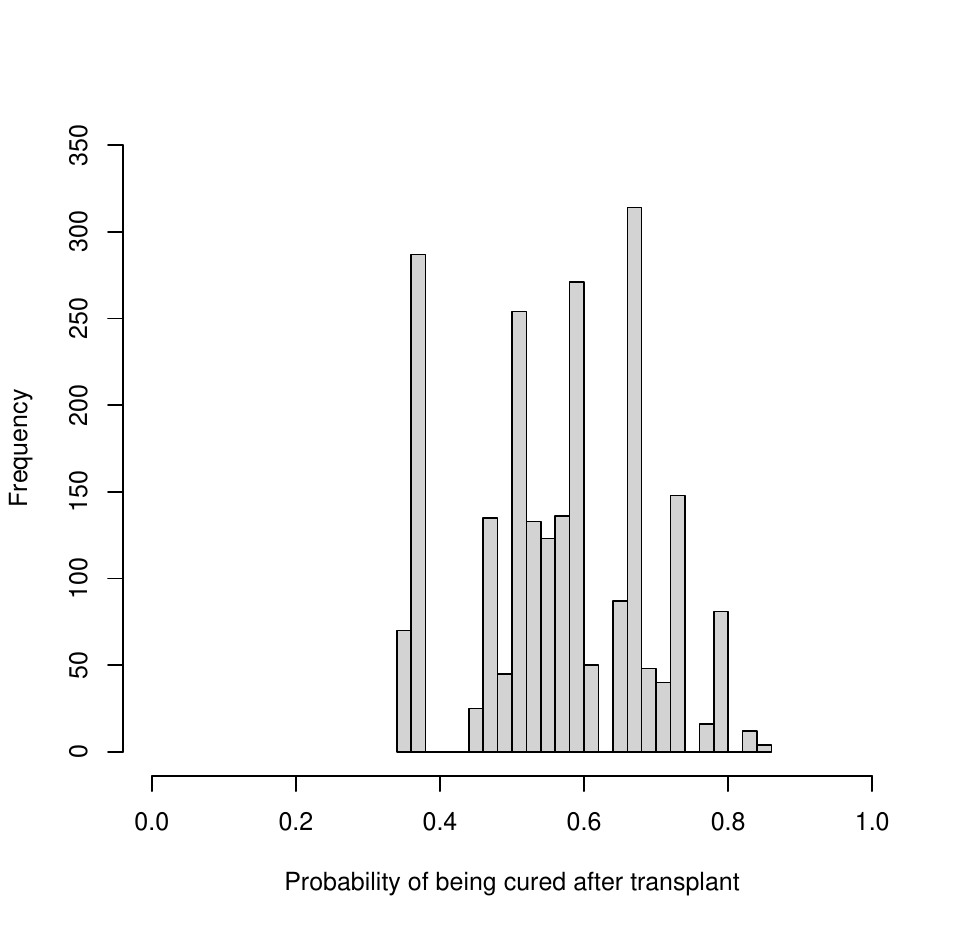}}
\caption{Probability of being cured based on cure analysis \label{prob of cure}}
\end{figure*}

For the 2279 patients from the EBMT, the probabilities of being cured after transplant are plotted in figure \ref{prob of cure}. Based on the baseline characteristics, the majority of patients are expected to be cured with a probability greater than 50\%.

\begin{landscape}
\begin{figure}%
\centerline{\includegraphics[width=1.5\textwidth]{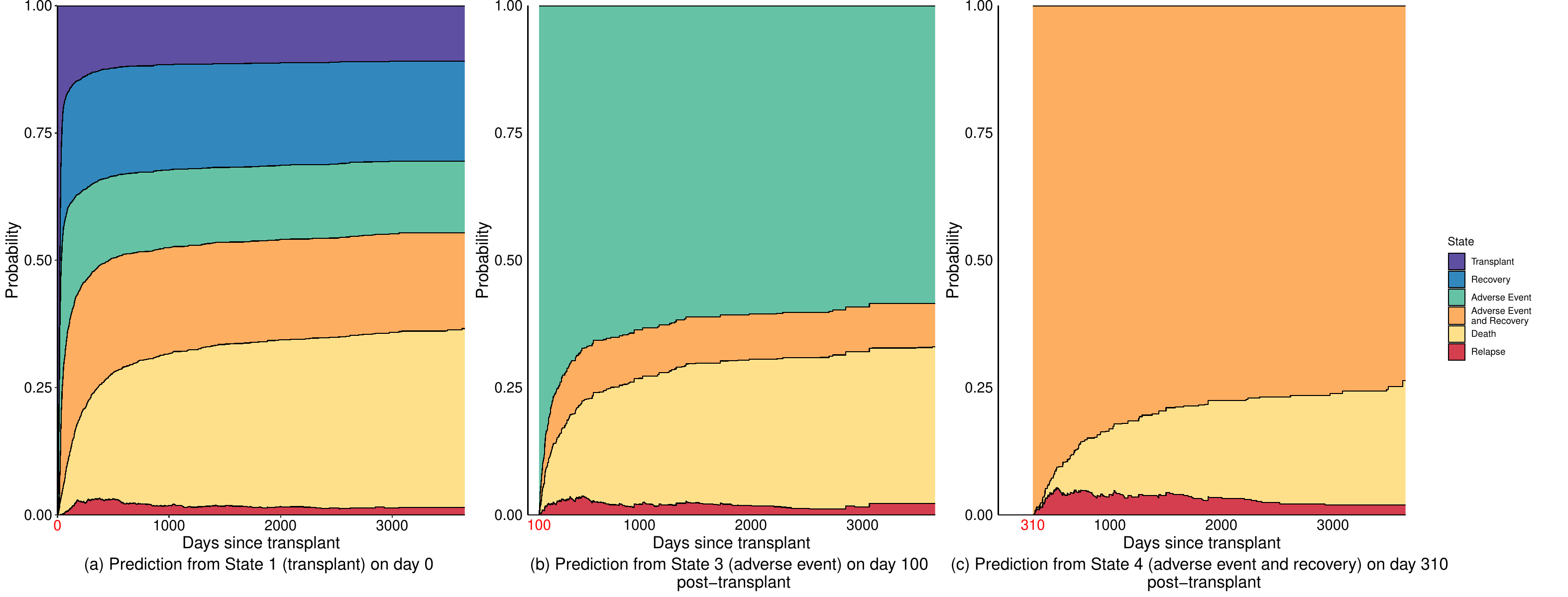}}
\caption{Stacked transition probabilities for dynamic prediction of subject 4 \label{fig:prediction}}
\end{figure}
\end{landscape}

For prediction, we use the subject in table \ref{new data wide} as an example. Subject 4 experienced an adverse event at 93 days after transplant, recovered at 310 days and was censored at 1925 days. Figure \ref{fig:prediction} presents the stacked transition probabilities for 10 years following the transplant, given different situations: (a) only the baseline covariates are known; (b)  an adverse event occurred on day 93 and the subject remained in the same state until day 100; (c)  the event history is known until day 310 where a recovery happened.
The state occupancy probabilities vary as more follow up information becomes available.

\section{Discussion}\label{sec7}

To date, no universal consensus exists regarding the definition of the multistate cure model framework. In this paper, we proposed a way of defining the multistate cure model through a set of non-cure states, $\mathbb{S}_{nc}$, where only subjects who transition to or from any of the non-cure states are known to be non-cured at baseline, while others have latent cure status.

To fit the multistate cure model, we developed an extended long data format that leverages the weighted likelihood presentation based on the EM algorithm. The method is highly flexible in model specification. With the EBMT data, we constructed a six-state Markov model including only baseline covariates. The cure component was defined through a single non-cure state, relapse. One may extend it to more complex scenarios, such as including additional states in general, integrating time-dependent covariates without assuming a Markov property, considering parametric baseline hazards or even allowing different baseline hazard functions for the same $k \rightarrow l$ transition but with different cure status. Additionaly, the set of non-cure states may include multiple states, e.g., subjects experiencing either local recurrence or  metastasis are non-cured. However, caution is needed, as increased model complexity and fewer restrictions may lead to numerical challenges.

Furthermore, our proposed method can facilitate dynamic prediction. As demonstrated, one can easily obtain the state occupancy probabilities within our multistate cure model framework, incorporating baseline covariates or even post-baseline information. With the aid of function \texttt{probtrans} from the \textbf{mstate} package, the exact formula does not need to be explicitly derived.

While defining the multistate cure model differently, in which cured and non-cured are two baseline states, Beesley and Taylor \cite{beesley2019algorithms} had a similar idea of decomposing the expected complete data log-likelihood $Q(\boldsymbol{\theta}|\boldsymbol{\theta}^{(r)})$ and making use of the EM algorithm. For the terms in $Q(\boldsymbol{\theta}|\boldsymbol{\theta}^{(r)})$ related to the survival analysis part, $Q_{1}(\boldsymbol{\beta},\boldsymbol{\gamma},\boldsymbol{\lambda_0(\cdot)}\,|\,\boldsymbol{\theta}^{(r)})$, Beesley and Taylor proposed a model-specific augmented data structure that contains rows for all model transitions including ones for which a subject is not at risk. In contrast, our extended long data structure, designed for generalizability, is built on top of the \texttt{msprep} function from the \textbf{mstate} package in R that involves only transitions for which a subject is at risk. For the terms related to the cure analysis, $Q_{2}(\alpha\,|\,\boldsymbol{\theta}^{(r)})$, Beesley and Taylor estimated $\alpha$ by fitting a logistic regression to $E_{\boldsymbol{\theta}^{(r)}}(G_{i}\,|\, \mathbb{D}_{i})$ using predictors $Z_{i,\alpha}$. In contrast, we keep utilizing an augmented data structure. As introduced in Section \ref{q2data}, each subject can have multiple rows with each row corresponding to a specific term in the $Q_{2}(\alpha\,|\,\boldsymbol{\theta}^{(r)})$ expression \eqref{Q2} that represents a baseline cure status, $G_i$, as indicated in column \texttt{cure}. A weighted logistic regression to $G_i$ is fit using this $Q_{2}(\alpha|\boldsymbol{\theta}^{(r)})$-focused extended 
 data format with $Z_{i,\alpha}$ being the predictors and $E_{\boldsymbol{\theta}^{(r)}}(G_{i}\,|\, \mathbb{D}_{i})$ or $1-E_{\boldsymbol{\theta}^{(r)}}(G_{i}\,|\, \mathbb{D}_{i})$ (depending on the value of $G_i$) being the weights, as indicated in column \texttt{weight}. The two different approaches give the same maximum-likelihood estimates when \texttt{cure} is binary, in which case the former one is easier to implement. However, the latter one, our approach, offers more flexibility and can be easily adapted for broader generalizations. One may expand the definition of cure or noncure to accommodate the population heterogeneity. For instance, when the set of non-cure states contain multiple states, different levels of non-cured statuses can be distinguished: patients who experienced local or regional recurrence can be classified as low-risk non-cured and patients who experienced distant recurrence as high-risk non-cured. In such a case,   the term $Q_{2}(\alpha|\boldsymbol{\theta}^{(r)})$ (see \eqref{Q2}), would resemble the log-likelihood for a weighted multinomial logistic regression. However, caution should be used with the definition to avoid potential identifiability issues. Beesley and Taylor \cite{beesley2019algorithms} also proposed a Monte Carlo EM (MCEM) algorithm to deal with covariate missingness and/or unequal follow-up of different time-to-event endpoints. The MCEM algorithm, originally introduced by Wei and Tanner \cite{wei1990monte}, is an imputation-based method designed to handle complex missing data within a modified EM algorithm. Further exploration of the MCEM algorithm could be valuable in the context of multistate cure models, particularly in scenarios with more complicated missing data mechanisms.

Due to the iterative nature, one disadvantage of the proposed algorithm is its relatively long running time. Sy and Taylor \cite{sy2000estimation} mentioned in their work on the standard Cox PH cure model that imposing a zero-tail constraint may accelerate convergence compared to the unconstraint MLE. We tested this by setting the weight $w_i^{(r)}$ to zero for those subjects who were censored for relapse after $t_{max}$ in the E-step, where $t_{max}$ is the last relapse time in the EBMT data. With a slightly lower overall observed data log-likelihood, the constrained MLE results conveyed the same message as what we had in table \ref{results}. The convergence was achieved faster with much less iterations. The improvement  was more visible in the bootstrap procedure.

Lastly, it is important to note that the application example with the EBMT data presented here is intended only for demonstration purposes and not for clinical interpretation, particularly as the issue of multiple testing was not addressed in the analysis.

The multistate cure model is particularly useful in pediatric oncology to study the complex courses of a disease where a proportion of the patient population achieves cure through treatment and therefore will not experience further progression or adverse events. It is a more realistic representation of disease dynamics where the patient population heterogeneity is well addressed. Our proposed EM algorithm, coupled with the extended long data format, supports model generalization and greatly simplifies the prediction process. It can provide valuable insights for prognosis prediction and assist in decision making and chronic disease management.

\section*{Acknowledgments}

A special thanks goes to Hein Putter for his invaluable help in shaping and formulating the ideas presented in this paper. 

\noindent The bootstrap procedure was performed on the supercomputer ALICE from Leiden University.

\section*{Conflict of interest}

The authors declare no potential conflict of interests.

\bibliographystyle{unsrt}  
\bibliography{references}

\section*{Supporting information}

The code is available at \url{https://github.com/yilin-jiang/cure-mstate-EM}.

\newpage
\begin{table}[h!]
\centering %
\caption{Patients baseline characteristics\label{baseline char}}
\resizebox{\columnwidth}{!}{%
\begin{tabular}{cccccccccc}
\hline
\multicolumn{3}{c}{\textbf{year of transplant}} &
  \multicolumn{3}{c}{\textbf{age}} &
  \multicolumn{2}{c}{\textbf{Prophylaxis}} &
  \multicolumn{2}{c}{\textbf{donor-recipient match}} \\
1985-1989 &
  1990-1994 &
  1995-1998 &
  \textless{}=20 &
  20-40 &
  \textgreater{}40 &
  no &
  yes &
  no gender mismatch &
  gender mismatch \\ \hline
634 &
  896 &
  749 &
  551 &
  1213 &
  515 &
  1730 &
  549 &
  1734 &
  545 \\ \hline
\end{tabular}
}
\end{table}

\begin{table}[h!]
\centering %
\caption{Example of EBMT data in wide format\label{data wide}}%
\begin{threeparttable}[t]
\begin{tabular*}{\textwidth}{
@{\extracolsep\fill}
p{0.2cm}p{0.4cm}p{0.4cm}p{0.4cm}p{0.4cm}p{0.4cm}p{0.6cm}p{0.4cm}p{0.4cm}p{0.4cm}
lllllllllllllll
@{\extracolsep\fill}
}
\toprule
\textbf{id} & \textbf{rec} & \textbf{rec.s} & \textbf{ae} & \textbf{ae.s} & \textbf{recae} & \textbf{recae.s} & \textbf{rel} & \textbf{rel.s} & \textbf{srv} & \textbf{srv.s} & \textbf{\begin{tabular}[c]{@{}l@{}}year of\\transplant\end{tabular}} & \textbf{age} & \textbf{\begin{tabular}[c]{@{}l@{}}prophy-\\ laxis\end{tabular}}& \textbf{\begin{tabular}[c]{@{}l@{}}donor-recipient\\ match\end{tabular}}\\
\midrule
1           & 143          & 0              & 60          & 1             & 143            & 0                & 143          & 0              & 143          & 1              & 1985-1989                   & \textgreater{}40           & yes                  & gender mismatch                \\
2           & 29           & 1              & 12          & 1             & 29             & 1                & 422          & 1              & 579          & 1              & 1995-1998                   & 20-40                      & no                   & no gender mismatch             \\
3 & 3687 & 0 & 14 & 1 & 3687 & 0 & 3687 & 0 & 3687 & 0 & 1990-1994 & 20-40 & no                   & no gender mismatch\\
\bottomrule
\end{tabular*}
\begin{tablenotes}
\item {\it Note}: Rec and rec.s indicate time in days from transplant to recovery or last follow-up and recovery status: 1 for recovery and 0 for censored. Ae and ae.s indicate Time in days from transplant to adverse event (AE) or last follow-up and AE status: 1 for AE and 0 for censored. Recae and Recae.s indicate time in days from transplant to both recovery and AE or last follow-up and the status: 1 for both recovery and AE, 0 for no recovery or no AE or censored. Rel and rel.s indicate time in days from transplant to relapse or last follow-up and relapse status: 1 for relapse and 0 for censored. Srv and srv.s indicate time in days from transplant to death or last follow-up and survival status: 1 for death and 0 for censored. Year of transplant, age at transplant, prophylaxis and donor-recipient gender match are the baseline covariates taken into account.
\end{tablenotes}
\end{threeparttable}
\end{table}

\begin{landscape}
\begin{table}[h]
\scriptsize
\caption{Example of EBMT data in extended long format (Q1-focused)\label{data long-long}}%
\begin{tabular*}{1.4\textwidth}
{p{0.1cm}p{0.28cm}p{0.2cm}p{0.4cm}p{0.4cm}p{0.4cm}p{0.4cm}p{0.65cm}p{0.6cm}p{0.4cm}p{0.4cm}p{0.5cm}p{0.5cm}p{0.5cm}p{0.5cm}p{0.5cm}p{0.45cm}p{0.45cm}p{0.45cm}p{0.45cm}
llllllllllllllllllllllll
}
\toprule
\textbf{id} & \textbf{from} & \textbf{to} & \textbf{trans} & \textbf{Tstart} & \textbf{Tstop}  & 
\textbf{status} &
\textbf{cumHaz} & \textbf{hazard} &
\textbf{\begin{tabular}[c]{@{}l@{}}likeli-\\ hood\end{tabular}} & \textbf{pi} &
\textbf{weight}   & \textbf{cure12} & \textbf{cure13} & \textbf{cure16} & \textbf{cure24} & \textbf{cure26} & \textbf{cure34} & \textbf{cure36} & \textbf{cure46} &  \textbf{\begin{tabular}[c]{@{}l@{}}year of\\transplant\end{tabular}}  & \textbf{age}     &  \textbf{\begin{tabular}[c]{@{}l@{}}prophy-\\ laxis\end{tabular}} &  \textbf{\begin{tabular}[c]{@{}l@{}}donor-recipient\\ match\end{tabular}}  \\
\midrule
\rowcolor{red!50}
1 & 1 & 2 & 1    & 0  & 60   & 0 & 0.329 &       & 0.720  & 0.830  & 0.123 & 0 & 0 & 0 & 0 & 0 & 0 & 0 & 0 & 1985-1989 & \textgreater{}40 & yes & gender mismatch    \\
\rowcolor{cyan}
1 & 1 & 2 & 1.2  & 0  & 60   & 0 & 0.379 &       & 0.685 & 0.830  & 0.877 & 1 & 0 & 0 & 0 & 0 & 0 & 0 & 0 & 1985-1989 & \textgreater{}40 & yes & gender mismatch    \\
\rowcolor{red!50}
1 & 1 & 3 & 2    & 0  & 60   & 1 & 0.344 & 0.004 & 0.003 & 0.830  & 0.123 & 0 & 0 & 0 & 0 & 0 & 0 & 0 & 0 & 1985-1989 & \textgreater{}40 & yes & gender mismatch    \\
\rowcolor{cyan}
1 & 1 & 3 & 2.2  & 0  & 60   & 1 & 0.489 & 0.006 & 0.004 & 0.830  & 0.877 & 0 & 1 & 0 & 0 & 0 & 0 & 0 & 0 & 1985-1989 & \textgreater{}40 & yes & gender mismatch    \\
\rowcolor{red!50}
1 & 1 & 5 & 3    & 0  & 60   & 0 & 0.037 &       & 0.964 & 0.830  & 0.123 & 0 & 0 & 0 & 0 & 0 & 0 & 0 & 0 & 1985-1989 & \textgreater{}40 & yes & gender mismatch    \\
\rowcolor{red!50}
1 & 1 & 6 & 4    & 0  & 60   & 0 & 0.112 &       & 0.894 & 0.830  & 0.123 & 0 & 0 & 0 & 0 & 0 & 0 & 0 & 0 & 1985-1989 & \textgreater{}40 & yes & gender mismatch    \\
\rowcolor{cyan}
1 & 1 & 6 & 4.2  & 0  & 60   & 0 & 0.14  &       & 0.869 & 0.830  & 0.877 & 0 & 0 & 1 & 0 & 0 & 0 & 0 & 0 & 1985-1989 & \textgreater{}40 & yes & gender mismatch    \\
\rowcolor{red!50}
1 & 3 & 4 & 8    & 60 & 143  & 0 & 0.092 &       & 0.912 & 0.830  & 0.123 & 0 & 0 & 0 & 0 & 0 & 0 & 0 & 0 & 1985-1989 & \textgreater{}40 & yes & gender mismatch    \\
\rowcolor{cyan}
1 & 3 & 4 & 8.2  & 60 & 143  & 0 & 0.104 &       & 0.901 & 0.830  & 0.877 & 0 & 0 & 0 & 0 & 0 & 1 & 0 & 0 & 1985-1989 & \textgreater{}40 & yes & gender mismatch    \\
\rowcolor{red!50}
1 & 3 & 5 & 9    & 60 & 143  & 0 & 0.339 &       & 0.713 & 0.830  & 0.123 & 0 & 0 & 0 & 0 & 0 & 0 & 0 & 0 & 1985-1989 & \textgreater{}40 & yes & gender mismatch    \\
\rowcolor{red!50}
1 & 3 & 6 & 10   & 60 & 143  & 1 & 0.527 & 0.007 & 0.004 & 0.830  & 0.123 & 0 & 0 & 0 & 0 & 0 & 0 & 0 & 0 & 1985-1989 & \textgreater{}40 & yes & gender mismatch    \\
\rowcolor{cyan}
1 & 3 & 6 & 10.2 & 60 & 143  & 1 & 0.429 & 0.006 & 0.004 & 0.830  & 0.877 & 0 & 0 & 0 & 0 & 0 & 0 & 1 & 0 & 1985-1989 & \textgreater{}40 & yes & gender mismatch    \\
\rowcolor{red!50}
2 & 1 & 2 & 1    & 0  & 12   & 0 & 0.009      &       &   0.991    &   0.509    & 1     & 0 & 0 & 0 & 0 & 0 & 0 & 0 & 0 & 1995-1998 & 20-40            & no  & no gender mismatch \\
\rowcolor{red!50}
2 & 1 & 3 & 2    & 0  & 12   & 1 & 0.093      &   0.018    &  0.017     &  0.509     & 1     & 0 & 0 & 0 & 0 & 0 & 0 & 0 & 0 & 1995-1998 & 20-40            & no  & no gender mismatch \\
\rowcolor{red!50}
2 & 1 & 5 & 3    & 0  & 12   & 0 &  0     &       &   1    & 0.509      & 1     & 0 & 0 & 0 & 0 & 0 & 0 & 0 & 0 & 1995-1998 & 20-40            & no  & no gender mismatch \\
\rowcolor{red!50}
2 & 1 & 6 & 4    & 0  & 12   & 0 &   0.006    &       &   0.994    &  0.509     & 1     & 0 & 0 & 0 & 0 & 0 & 0 & 0 & 0 & 1995-1998 & 20-40            & no  & no gender mismatch \\
\rowcolor{red!50}
2 & 3 & 4 & 8    & 12 & 29   & 1 &  0.413     &  0.030     &  0.020     & 0.509      & 1     & 0 & 0 & 0 & 0 & 0 & 0 & 0 & 0 & 1995-1998 & 20-40            & no  & no gender mismatch \\
\rowcolor{red!50}
2 & 3 & 5 & 9    & 12 & 29   & 0 & 0.003      &       & 0.997      &  0.509     & 1     & 0 & 0 & 0 & 0 & 0 & 0 & 0 & 0 & 1995-1998 & 20-40            & no  & no gender mismatch \\
\rowcolor{red!50}
2 & 3 & 6 & 10   & 12 & 29   & 0 &  0.025     &       &   0.976    &  0.509     & 1     & 0 & 0 & 0 & 0 & 0 & 0 & 0 & 0 & 1995-1998 & 20-40            & no  & no gender mismatch \\
\rowcolor{red!50}
2 & 4 & 5 & 11   & 29 & 422  & 1 & 0.367      &  0.008     &    0.006   & 0.509      & 1     & 0 & 0 & 0 & 0 & 0 & 0 & 0 & 0 & 1995-1998 & 20-40            & no  & no gender mismatch \\
\rowcolor{red!50}
2 & 4 & 6 & 12   & 29 & 422  & 0 &  0.238     &       &0.788       & 0.509      & 1     & 0 & 0 & 0 & 0 & 0 & 0 & 0 & 0 & 1995-1998 & 20-40            & no  & no gender mismatch \\
\rowcolor{red!50}
2 & 5 & 6 & 13   & 422 & 579  & 1 &  0.727     &  0.056     & 0.027      & 0.509      & 1     & 0 & 0 & 0 & 0 & 0 & 0 & 0 & 0 & 1995-1998 & 20-40            & no  & no gender mismatch \\
\rowcolor{red!50}
3 & 1 & 2 & 1    & 0  & 0 & 14   & 0.025 &       & 0.975 & 0.378 & 0.479 & 0 & 0 & 0 & 0 & 0 & 0 & 0 & 0 & 1990-1994 & 20-40            & no  & no gender mismatch \\
\rowcolor{cyan}
3 & 1 & 2 & 1.2  & 0  & 0 & 14   & 0.029 &       & 0.971 & 0.378 & 0.521 & 1 & 0 & 0 & 0 & 0 & 0 & 0 & 0 & 1990-1994 & 20-40            & no  & no gender mismatch \\
\rowcolor{red!50}
3 & 1 & 3 & 2    & 0  & 1 & 14   & 0.176 & 0.033 & 0.027 & 0.378 & 0.479 & 0 & 0 & 0 & 0 & 0 & 0 & 0 & 0 & 1990-1994 & 20-40            & no  & no gender mismatch \\
\rowcolor{cyan}
3 & 1 & 3 & 2.2  & 0  & 1 & 14   & 0.249 & 0.046 & 0.036 & 0.378 & 0.521 & 0 & 1 & 0 & 0 & 0 & 0 & 0 & 0 & 1990-1994 & 20-40            & no  & no gender mismatch \\
\rowcolor{red!50}
3 & 1 & 5 & 3    & 0  & 0 & 14   & 0     &       & 1     & 0.378 & 0.479 & 0 & 0 & 0 & 0 & 0 & 0 & 0 & 0 & 1990-1994 & 20-40            & no  & no gender mismatch \\
\rowcolor{red!50}
3 & 1 & 6 & 4    & 0  & 0 & 14   & 0.009 &       & 0.991 & 0.378 & 0.479 & 0 & 0 & 0 & 0 & 0 & 0 & 0 & 0 & 1990-1994 & 20-40            & no  & no gender mismatch \\
\rowcolor{cyan}
3 & 1 & 6 & 4.2  & 0  & 0 & 14   & 0.011 &       & 0.989 & 0.378 & 0.521 & 0 & 0 & 1 & 0 & 0 & 0 & 0 & 0 & 1990-1994 & 20-40            & no  & no gender mismatch \\
\rowcolor{red!50}
3 & 3 & 4 & 8    & 14 & 0 & 3687 & 0.697 &       & 0.498 & 0.378 & 0.479 & 0 & 0 & 0 & 0 & 0 & 0 & 0 & 0 & 1990-1994 & 20-40            & no  & no gender mismatch \\
\rowcolor{cyan}
3 & 3 & 4 & 8.2  & 14 & 0 & 3687 & 0.791 &       & 0.453 & 0.378 & 0.521 & 0 & 0 & 0 & 0 & 0 & 1 & 0 & 0 & 1990-1994 & 20-40            & no  & no gender mismatch \\
\rowcolor{red!50}
3 & 3 & 5 & 9    & 14 & 0 & 3687 & 0.339 &       & 0.713 & 0.378 & 0.479 & 0 & 0 & 0 & 0 & 0 & 0 & 0 & 0 & 1990-1994 & 20-40            & no  & no gender mismatch \\
\rowcolor{red!50}
3 & 3 & 6 & 10   & 14 & 0 & 3687 & 0.364 &       & 0.695 & 0.378 & 0.479 & 0 & 0 & 0 & 0 & 0 & 0 & 0 & 0 & 1990-1994 & 20-40            & no  & no gender mismatch \\
\rowcolor{cyan}
3 & 3 & 6 & 10.2 & 14 & 0 & 3687 & 0.297 &       & 0.743 & 0.378 & 0.521 & 0 & 0 & 0 & 0 & 0 & 0 & 1 & 0 & 1990-1994 & 20-40            & no  & no gender mismatch \\
\bottomrule
\end{tabular*}
\begin{tablenotes}
\item {\it Note}: Rows corresponding to non-cured status are in red. Rows corresponding to cured status are in blue.
\end{tablenotes}
\end{table}
\end{landscape}

\begin{table}[h!]
\caption{Example of EBMT data (Q2-focused)\label{data q2 focused}}%
\begin{tabular*}{\textwidth}{@{\extracolsep\fill}lllllll@{\extracolsep\fill}}
\toprule
\textbf{id} & \textbf{weight} & \textbf{cure} & \textbf{year of transplant} & \textbf{age} & \textbf{prophylaxis} & \textbf{donor-recipient match} \\
\midrule
1 & 0.123 & 0 & 1985-1989 & \textgreater{}40 & yes & gender mismatch    \\
1 & 0.877 & 1 & 1985-1989 & \textgreater{}40 & yes & gender mismatch    \\
2  & 1    & 0 & 1995-1998 & 20-40            & no  & no gender mismatch \\
3  & 0.479    & 0 & 1990-1994 & 20-40            & no  & no gender mismatch\\
3  & 0.521    & 1 & 1990-1994 & 20-40            & no  & no gender mismatch\\
\bottomrule
\end{tabular*}
\end{table}

\begin{landscape}
\begin{table}[h]
\caption{Regression coefficients (standard errors)\label{results}}%
\resizebox{\textwidth}{!}{%
\begin{tabular}{llllllll}
\hline
cure analysis &              & \multicolumn{2}{c}{Year of transplant}            & \multicolumn{2}{c}{Age at transplant} & Prophylaxis             & donor-recepient  \\
&
  intercept &
  \multicolumn{1}{c}{1990-1994} &
  \multicolumn{1}{c}{1995-1998} &
  \multicolumn{1}{c}{20-40} &
  \multicolumn{1}{c}{\textgreater{}40} &yes
   & gender mismatch
   \\ \hline
   & 0.669 (0.521) & -0.824 (0.515)           & -0.29 (0.563)            & -0.345 (0.436)     & 0.359 (0.560)    & -0.081 (0.451)           & 0.638 (0.420)         \\ \hline
  survival analysis &             & \multicolumn{2}{c}{Year of transplant}            & \multicolumn{2}{c}{Age at transplant} & Prophylaxis             & donor-recepient  \\
transition &
  cure &
  \multicolumn{1}{c}{1990-1994} &
  \multicolumn{1}{c}{1995-1998} &
  \multicolumn{1}{c}{20-40} &
  \multicolumn{1}{c}{\textgreater{}40} & yes
   & gender mismatch
   \\ \hline
1$\rightarrow$2 &
  0.141 (0.487) &
  \textbf{0.430 (0.151)} &
  \textbf{0.533 (0.134)} &
  0.060 (0.107) &
  0.185 (0.138) &
  \textbf{-0.365 (0.102)} &
  -0.191 (0.128) \\
1$\rightarrow$3 & 0.350 (0.362)  & 0.091 (0.137)           & -0.089 (0.125)          & 0.151 (0.103)     & 0.033 (0.133)     & \textbf{-0.274 (0.103)} & -0.165 (0.114)         \\
1$\rightarrow$5 &                & -0.055 (0.503)          & 0.062 (0.610)           & -0.279 (0.475)    & 0.082 (0.743)     & 0.472 (0.429)           & 0.65 (0.542)           \\
1$\rightarrow$6 &
  0.224 (1.701) &
  -0.314 (0.376) &
  -0.461 (0.280) &
  \textbf{0.787 (0.286)} &
  \textbf{0.914 (0.354)} &
  -0.059 (0.247) &
  -0.043 (0.333) \\
2$\rightarrow$4 & 0.400 (0.826)  & -0.015 (0.312)          & -0.126 (0.263)          & 0.321 (0.224)     & 0.438 (0.274)     & -0.279 (0.251)          & 0.132 (0.253)          \\
2$\rightarrow$5 &                & -0.795 (0.460)          & -0.102 (0.527)          & -0.681 (0.451)    & 0.071 (0.481)     & 0.325 (0.432)           & \textbf{1.067 (0.436)} \\
2$\rightarrow$6 &                & \textbf{-0.836 (0.392)} & \textbf{-0.980 (0.439)} & 0.150 (0.980)     & 1.465 (0.959)     & -0.008 (0.387)          & 0.244 (0.731)          \\
3$\rightarrow$4 &
  0.127 (0.509) &
  \textbf{0.553 (0.193)} &
  \textbf{0.937 (0.161)} &
  \textbf{-0.383 (0.131)} &
  \textbf{-0.333 (0.168)} &
  0.126 (0.136) &
  0.109 (0.143) \\
3$\rightarrow$5 &                & -1.197 (0.648)          & -1.091 (1.207)          & -0.027 (0.547)    & 0.873 (0.892)     & 0.077 (0.707)           & 0.107 (0.647)          \\
3$\rightarrow$6 & -0.205 (1.226) & \textbf{-0.69 (0.303)}  & -0.232 (0.246)          & 0.225 (0.250)     & 0.515 (0.299)     & 0.324 (0.204)           & 0.038 (0.232)          \\
4$\rightarrow$5 &                & \textbf{-1.709 (0.626)} & \textbf{-1.545 (0.717)} & 0.154 (0.469)     & 0.888 (0.834)     & 0.062 (0.554)           & 0.618 (0.685)          \\
4$\rightarrow$6 &
  -1.002 (1.557) &
  -0.659 (0.484) &
  -0.505 (0.352) &
  \textbf{0.696 (0.309)} &
  \textbf{1.464 (0.376)} &
  -0.129 (0.293) &
  \textbf{0.729 (0.280)} \\
  5$\rightarrow$6 &
  & -0.256 (0.146) &
  \textbf{-0.360 (0.162)} &
  \textbf{0.357 (0.144)} &
  0.165 (0.179) &
  -0.006 (0.130) &
  0.038 (0.138)\\
  \hline
\end{tabular}%
}
\begin{tablenotes}
\item {\it Note}: Covariate effects significant at 0.05 are in bold. In the survival analysis part, the effect of baseline cure status, $\gamma$, is included in the column \texttt{cure}. 
\end{tablenotes}
\end{table}

\begin{table}[h!]
\centering %
\caption{Subject information for prediction\label{new data wide}}%
\begin{tabular*}{1.4\textwidth}{@{\extracolsep\fill}p{0.2cm}p{0.4cm}p{0.4cm}p{0.4cm}p{0.4cm}p{0.4cm}p{0.6cm}p{0.4cm}p{0.4cm}p{0.4cm}lllllllllllllll@{\extracolsep\fill}}
\toprule
\textbf{id} & \textbf{rec} & \textbf{rec.s} & \textbf{ae} & \textbf{ae.s} & \textbf{recae} & \textbf{recae.s} & \textbf{rel} & \textbf{rel.s} & \textbf{srv} & \textbf{srv.s} & \textbf{year of transplant} & \textbf{age} & \textbf{prophylaxis} & \textbf{donor-recipient match} \\
\midrule
4           & 310          & 1              & 93          & 1             & 310            & 1                & 1925          & 0              & 1925        & 0             & 1990-1994                  & 20-40          & no                  & no gender mismatch         \\
\bottomrule
\end{tabular*}
\end{table}
\end{landscape}
\newpage

\section*{Appendix}

\section*{Calculation of the observed information matrix}\label{app1}

Unlike in the standard Cox model, the submatrix of $I(\alpha, \boldsymbol{\beta},\boldsymbol{\gamma},\boldsymbol{\lambda_0(\cdot)})$ corresponding to $\boldsymbol{\lambda_0(\cdot)}$ is not diagonal, therefore the asymptotic variance of $(\alpha, \boldsymbol{\beta},\boldsymbol{\gamma})$ is based on the full observed information matrix.

Recall the EM-operator from \eqref{expected com likelihood},\eqref{Q1},\eqref{Q2},\eqref{w}, given by
\begin{equation}\label{EM-operator}
\scriptsize
    Q(\boldsymbol{\theta}|\boldsymbol{\theta}^{(r)})=\sum_{i=1}^{n}\Bigl\{ w_{i}^{(r)}\,\mathrm{log}P_{\boldsymbol{\theta}}(\mathbb{D}_{i} \,|\, G_{i}=1, Z_{i,\beta})+(1-w_{i}^{(r)})\,\mathrm{log}P_{\boldsymbol{\theta}}(\mathbb{D}_{i}\, |\, G_{i}=0, Z_{i,\beta})+\\
     w_{i}^{(r)}\,\mathrm{log}P_{\boldsymbol{\theta}}(G_{i}=1 \,|\,  Z_{i,\alpha})+(1-w_{i}^{(r)})\,\mathrm{log}P_{\boldsymbol{\theta}}(G_{i}=0\, |\, Z_{i,\alpha})\Bigr\}.
\end{equation}

The method of Oakes\cite{oakes1999direct} shows that the observed information of the maximum likelihood estimator from the EM algorithm can be obtained via 
\begin{equation}\label{observed info formula}
    -\, \frac{\partial^2 \ell_{O}(\boldsymbol{\theta})}{\partial \boldsymbol{\theta}^2}= -\, \frac{\partial^2 Q(\boldsymbol{\theta}|\boldsymbol{\theta}^{(r)})}{\partial\boldsymbol{\theta}\partial\boldsymbol{\theta}^T}\Big|_{\boldsymbol{\theta}=\boldsymbol{\hat{\theta}}} \,- \, \frac{\partial^2 Q(\boldsymbol{\theta}|\boldsymbol{\theta}^{(r)})}{\partial\boldsymbol{\theta}\partial\boldsymbol{\theta}^{(r)T}}\Big|_{\boldsymbol{\theta}=\boldsymbol{\hat{\theta}}}.
\end{equation}

Analogous to obtaining the observed information matrix from a standard logistic regression, 
\begin{equation}\label{standard logit score}
    \frac{\partial Q(\boldsymbol{\theta}|\boldsymbol{\theta}^{(r)})}{\partial\alpha}=\sum_{i=1}^{n}\bigl[w_i^{(r)} - \frac{\exp(Z_{i,\alpha}^T \alpha)}{1+\exp(Z_{i,\alpha}^T \alpha)}\bigr]Z_{i,\alpha},
\end{equation}

\begin{equation}\label{standard logit hessian}
    -\,\frac{\partial^2 Q(\boldsymbol{\theta}|\boldsymbol{\theta}^{(r)})}{\partial\alpha\partial\alpha^T}=\sum_{i=1}^{n}\frac{\exp(Z_{i,\alpha}^T \alpha)}{\bigl(1+\exp(Z_{i,\alpha}^T \alpha)\bigr)^2}Z_{i,\alpha}Z_{i,\alpha}^T.
\end{equation}

Estimate for $ -\,\frac{\partial^2 Q(\boldsymbol{\theta}|\boldsymbol{\theta}^{(r)})}{\partial\alpha\partial\alpha^T}\Big|_{\alpha=\hat{\alpha}}$ can be obtained taking the inverse of the covariance results from the weighted logistic model fit at the last iteration. \\

For simplicity, consider $\boldsymbol{\beta}$ and $\boldsymbol{\gamma}$ together as $\boldsymbol{b}=\{\boldsymbol{\beta},\boldsymbol{\gamma}\}$ and let $Z_b=\{Z_{\beta}, G\}$. We make use of the Breslow's estimator \cite{breslow1972contribution} for the baseline cumulative hazard $\boldsymbol{\Lambda_0(\cdot)}$ where $\boldsymbol{\hat{\Lambda}_0(\cdot)}$ is piecewise constant between event times. Suppose for transition $k\rightarrow l$ we have $m_{kl}$ distinct event times and we denote them by $t_{(1)} < \cdots < t_{(m_{kl})}$, therefore with $t_{(0)}=0$ we have,
\begin{equation}\label{baseline hazard at event times}
\hat{\lambda}_{0,j,kl}=\hat{\Lambda}_{0,kl}(t_{(j)})-\hat{\Lambda}_{0,kl}(t_{(j-1)}), \qquad j=1,...,m_{kl},
\end{equation}
in which the baseline hazard function is estimated as the incremental change in the baseline cumulative hazard at event times.
Treating $\lambda_{0,j,kl}$ as parameters, we can plug the Breslow's estimates for the baseline hazard function into \eqref{likelihood per subject given G} and rewrite the formula as 
\begin{equation}\label{rewritten likelihood}
    P_{\boldsymbol{\theta}}(\mathbb{D}_{i} \,|\,Z_{i,b}) =\prod_{k=1}^{s}\prod_{l \neq k=1}^{s} \Bigl\{ \prod_{t_{(j)}=0}^{X_i}
\bigl[\lambda_{0,j,kl}\exp(Z_{i,b}^T b_{kl}) ]^{dN_{i,kl}(t_{(j)})}\Bigr\}\,\mathrm{exp}\Bigl(-\sum_{t_{(j)}=0}^{X_i}Y_{i,k}(t_{(j)})\lambda_{0,j,kl}\exp(Z_{i,b}^T b_{kl}) \Bigr).
\end{equation}

Taking the logarithm of \eqref{rewritten likelihood} yields 

\begin{equation}\label{standard cox}
\mathrm{log}P_{\boldsymbol{\theta}}(\mathbb{D}_{i} \,|\, Z_{i,b})=\sum_{k=1}^{s}\sum_{l \neq k=1}^{s}\sum_{j:t_{(j)}\leq X_i}\bigl\{ [\mathrm{log}\lambda_{0,j,kl}+Z_{i,b}^T b_{kl}]dN_{i,kl}(t_{(j)})-Y_{i,k}(t_{(j)})\lambda_{0,j,kl}\exp(Z_{i,b}^T b_{kl}) \bigr\}. 
\end{equation}

For $k\rightarrow l$ transition, we obtain 
\begin{equation} \label{standard cox score for beta}
 \frac{\partial \mathrm{log}P_{\boldsymbol{\theta}}(\mathbb{D}_{i} \,|\, Z_{i,b})}{\partial b_{kl}}\,=\,\sum_{j:t_{(j)}\leq X_i} \bigl\{Z_{i,b}dN_{i,kl}(t_{(j)})-Y_{i,k}(t_{(j)})\lambda_{0,j,kl}Z_{i,b}^T\exp(Z_{i,b}^T b_{kl})\bigr\},
\end{equation}
\begin{equation}\label{standard cox score for hazard}
     \frac{\partial \mathrm{log}P_{\boldsymbol{\theta}}(\mathbb{D}_{i} \,|\, Z_{i,b})}{\partial \lambda_{0,j,kl}}\,=\frac{dN_{i,kl}(t_{(j)})}{\lambda_{0,j,kl}}-Y_{i,k}(t_{(j)})\exp(Z_{i,b}^T b_{kl}).
\end{equation}

therefore,
\begin{equation}\label{first derivative b}
\frac{\partial Q(\boldsymbol{\theta}|\boldsymbol{\theta}^{(r)})}{\partial b_{kl}}\,=\,\sum_{i=1}^{n} \sum_{G_i=0}^{1}\sum_{j:t_{(j)}\leq X_i}\bigl[G_{i} w_i^{(r)}+ (1-G_i)(1-w_i^{(r)})\bigr]\bigl\{Z_{i,b}dN_{i,kl}(t_{(j)})-Y_{i,k}(t_{(j)})\lambda_{0,j,kl}Z_{i,b}^T\exp(Z_{i,b}^T b_{kl})\bigr\},
\end{equation}

\begin{equation}\label{first derivative hazard}
    \frac{\partial Q(\boldsymbol{\theta}|\boldsymbol{\theta}^{(r)})}{\partial \lambda_{0,j,kl}}\,=\,\sum_{i=1}^{n} \sum_{G_i=0}^{1}\bigl[G_{i} w_i^{(r)}+ (1-G_i)(1-w_i^{(r)})\bigr]\bigl\{\frac{dN_{i,kl}(t_{(j)})}{\lambda_{0,j,kl}}-Y_{i,k}(t_{(j)})\exp(Z_{i,b}^T b_{kl})\bigr\}.
\end{equation}

The components of $-\frac{\partial^2 Q(\boldsymbol{\theta}|\boldsymbol{\theta}^{(r)})}{\partial\boldsymbol{\theta}\partial\boldsymbol{\theta}^T}\Big|_{\boldsymbol{\theta}=\boldsymbol{\hat{\theta}}}$with respect to $\{\boldsymbol{b},\boldsymbol{\lambda_0(\cdot)}\}$ are
\begin{equation}\label{first component b2}
-\frac{\partial^2 Q(\boldsymbol{\theta}|\boldsymbol{\theta}^{(r)})}{\partial b_{kl}\partial b_{kl}^T}\,=\,\sum_{i=1}^{n} \sum_{G_i=0}^{1}\sum_{j:t_{(j)}\leq X_i}\bigl[G_{i} w_i^{(r)}+ (1-G_i)(1-w_i^{(r)})\bigr]Y_{i,k}(t_{(j)})\lambda_{0,j,kl}Z_{i,b}Z_{i,b}^T\exp(Z_{i,b}^T b_{kl}),
\end{equation}

\begin{equation}\label{first component lambda2}
-\frac{\partial^2 Q(\boldsymbol{\theta}|\boldsymbol{\theta}^{(r)})}{\partial \lambda_{0,j,kl}^2}\,=\,\sum_{i=1}^{n} \sum_{G_i=0}^{1}\bigl[G_{i} w_i^{(r)}+ (1-G_i)(1-w_i^{(r)})\bigr]\frac{dN_{i,kl}(t_{(j)})}{\lambda_{0,j,kl}^2},
\end{equation}

\begin{equation}\label{first component b lambda}
-\frac{\partial^2 Q(\boldsymbol{\theta}|\boldsymbol{\theta}^{(r)})}{\partial b_{kl}\partial\lambda_{0,j,kl}}\,=\,\sum_{i=1}^{n} \sum_{G_i=0}^{1}\sum_{j:t_{(j)}\leq X_i}\bigl[G_{i} w_i^{(r)}+ (1-G_i)(1-w_i^{(r)})\bigr]Y_{i,k}(t_{(j)})Z_{i,b}^T\exp(Z_{i,b}^T b_{kl}),
\end{equation}

\begin{equation}\label{first component b b}
-\frac{\partial^2 Q(\boldsymbol{\theta}|\boldsymbol{\theta}^{(r)})}{\partial b_{kl}\partial b_{k'l'}}\,=0, \qquad unless \quad k=k' \quad and \quad l=l',
\end{equation}

\begin{equation}\label{first component lambda lambda}
-\frac{\partial^2 Q(\boldsymbol{\theta}|\boldsymbol{\theta}^{(r)})}{\partial \lambda_{0,j,kl}\partial \lambda_{0,j',k'l'}^T}\,=0, \qquad unless \quad j=j' \quad and \quad k=k' \quad and \quad l=l',
\end{equation}

\noindent together with \eqref{standard logit hessian},  now the first term in \eqref{observed info formula}, $ -\, \frac{\partial^2 Q(\boldsymbol{\theta}|\boldsymbol{\theta}^{(r)})}{\partial\boldsymbol{\theta}\partial\boldsymbol{\theta}^T}\Big|_{\boldsymbol{\theta}=\boldsymbol{\hat{\theta}}}\,$, has been solved.  Note that one can utilize the \texttt{basehaz} function in the \textbf{survival} package and then take the incremental change to obtain $\hat{\lambda}_{0,j,kl}$.
One may also take the inverse of the covariance results from the weighted Cox model fit for transition $k\rightarrow l$ at the last iteration to get the direct estimates for \eqref{first component b2}. 

 For the second term, $-\frac{\partial^2 Q(\boldsymbol{\theta}|\boldsymbol{\theta}^{(r)})}{\partial\boldsymbol{\theta}\partial\boldsymbol{\theta}^{(r)T}}$, we still need to derive $\frac{\partial w_i^{(r)}}{\partial \boldsymbol{\theta^{(r)}}}$.\\

Define shorthand $\boldsymbol{\phi}=\{\boldsymbol{b},\boldsymbol{\lambda_0(\cdot)}\}$. Since we have 
\begin{equation}\label{chain1}
\begin{split}
   \frac{\partial w_i^{(r)}}{\partial \pi_i^{(r)}}& =\frac{\partial}{\partial \pi_i^{(r)}}\frac{\pi_{i}^{(r)}P_{\boldsymbol{\phi}^{(r)}}\bigl(\mathbb{D}_{i} \,|\, G_{i}=1, Z_{i,\beta}\bigr)}{\pi_{i}^{(r)}P_{\boldsymbol{\phi}^{(r)}}\bigl(\mathbb{D}_{i} \,|\, G_{i}=1, Z_{i,\beta}\bigr)+(1-\pi_{i}^{(r)})P_{\boldsymbol{\phi}^{(r)}}\bigl(\mathbb{D}_{i} \,|\, G_{i}=0, Z_{i,\beta}\bigr)}\\
   &=\frac{P_{\boldsymbol{\phi}^{(r)}}\bigl(\mathbb{D}_{i} \,|\, G_{i}=0, Z_{i,\beta}\bigr)P_{\boldsymbol{\phi}^{(r)}}\bigl(\mathbb{D}_{i} \,|\, G_{i}=1, Z_{i,\beta}\bigr)}{\Bigl[\pi_{i}^{(r)}P_{\boldsymbol{\phi}^{(r)}}\bigl(\mathbb{D}_{i} \,|\, G_{i}=1, Z_{i,\beta}\bigr)+(1-\pi_{i}^{(r)})P_{\boldsymbol{\phi}^{(r)}}\bigl(\mathbb{D}_{i} \,|\, G_{i}=0, Z_{i,\beta}\bigr)  \Bigr]^2}
\end{split}
\end{equation}
and
\begin{equation}\label{chain2}
\begin{split}
    \frac{\partial \pi_i^{(r)}}{\partial \alpha^{(r)}}&=\frac{\partial }{\partial \alpha^{(r)}}P_{\alpha^{(r)}}(G_{i}=1\,|\,Z_{i,\alpha})\\
    &=\frac{\partial }{\partial \alpha^{(r)}}\frac{1}{1+\exp(-Z_{i,\alpha}^T\alpha^{(r)})}\\
    &=\frac{Z_{i,\alpha}^T\exp(-Z_{i,\alpha}^T\alpha^{(r)})}{\bigl[1+\exp(-Z_{i,\alpha}^T\alpha^{(r)})\bigr]^2},
\end{split}
\end{equation}
therefore, based on the chain rule,
\begin{equation}\label{chain rule}
\begin{split}
    \frac{\partial w_i^{(r)}}{\partial \alpha^{(r)}}&=\frac{\partial w_i^{(r)}}{\partial \pi_i^{(r)}}\cdot\frac{\partial \pi_i^{(r)}}{\partial \alpha^{(r)}}\\
    &=\frac{P_{\boldsymbol{\phi}^{(r)}}\bigl(\mathbb{D}_{i} \,|\, G_{i}=0, Z_{i,\beta}\bigr)P_{\boldsymbol{\phi}^{(r)}}\bigl(\mathbb{D}_{i} \,|\, G_{i}=1, Z_{i,\beta}\bigr)Z_{i,\alpha}^T\exp(-Z_{i,\alpha}^T\alpha^{(r)})}{\bigl[P_{\boldsymbol{\phi}^{(r)}}\bigl(\mathbb{D}_{i} \,|\, G_{i}=1, Z_{i,\beta}\bigr)-\exp(-Z_{i,\alpha}^T\alpha^{(r)})P_{\boldsymbol{\phi}^{(r)}}\bigl(\mathbb{D}_{i} \,|\, G_{i}=0, Z_{i,\beta}\bigr)  \bigr]^2},
\end{split}
\end{equation}
where $P_{\boldsymbol{\phi}^{(r)}}\bigl(\mathbb{D}_{i} \,|\, G_{i}, Z_{i,\beta}\bigr)$ is calculated at $\boldsymbol{\phi}^{(r)}$ using formula \eqref{likelihood per subject given G}.\\

With \eqref{chain rule}, \eqref{standard logit score},\eqref{first derivative b} and \eqref{first derivative hazard}, we can obtain $-\frac{\partial^2 Q(\boldsymbol{\theta}|\boldsymbol{\theta}^{(r)})}{\partial\alpha\partial\alpha^{(r)T}}$, $-\frac{\partial^2 Q(\boldsymbol{\theta}|\boldsymbol{\theta}^{(r)})}{\partial b_{kl}\partial\alpha^{(r)T}}$ and $-\frac{\partial^2 Q(\boldsymbol{\theta}|\boldsymbol{\theta}^{(r)})}{\partial\lambda_{0,j,kl}\partial\alpha^{(r)T}}$ via the following relationships. The lengthy expressions are omitted here.

\begin{equation}\label{second term alpha alpha r}
    -\frac{\partial^2 Q(\boldsymbol{\theta}|\boldsymbol{\theta}^{(r)})}{\partial\alpha\partial\alpha^{(r)T}} = -\sum_{i=1}^{n}\frac{\partial w_i^{(r)}}{\partial \alpha^{(r)}}Z_{i,\alpha},
\end{equation}
\begin{equation}\label{second term b alpha r}
-\frac{\partial^2 Q(\boldsymbol{\theta}|\boldsymbol{\theta}^{(r)})}{\partial b_{kl}\partial\alpha^{(r)T}} =  -\sum_{i=1}^{n} \sum_{G_i=0}^{1}\sum_{j:t_{(j)}\leq X_i}  (2G_{i}-1)\frac{\partial w_i^{(r)}}{\partial \alpha^{(r)}} \bigl\{Z_{i,b}dN_{i,kl}(t_{(j)})-Y_{i,k}(t_{(j)})\lambda_{0,j,kl}Z_{i,b}^T\exp(Z_{i,b}^T b_{kl})\bigr\} ,
\end{equation}

\begin{equation}\label{second term lambda alpha r}
 -\frac{\partial^2 Q(\boldsymbol{\theta}|\boldsymbol{\theta}^{(r)})}{\partial\lambda_{0,j,kl}\partial\alpha^{(r)T}} =  - \sum_{i=1}^{n} \sum_{G_i=0}^{1}(2G_{i}-1)\frac{\partial w_i^{(r)}}{\partial \alpha^{(r)}}\bigl\{\frac{dN_{i,kl}(t_{(j)})}{\lambda_{0,j,kl}}-Y_{i,k}(t_{(j)})\exp(Z_{i,b}^T b_{kl})\bigr\}. 
\end{equation}

\noindent Define shorthand $f(\boldsymbol{\phi}^{(r)})=P_{\boldsymbol{\phi}^{(r)}}\bigl(\mathbb{D}_{i} \,|\, G_{i}=1, Z_{i,\beta}\bigr)$ and $h(\boldsymbol{\phi}^{(r)})=P_{\boldsymbol{\phi}^{(r)}}\bigl(\mathbb{D}_{i} \,|\, G_{i}=0, Z_{i,\beta}\bigr)$.

\noindent Note that 
\begin{align*} 
f'(\boldsymbol{\phi}^{(r)})=\frac{\partial \mathrm{log} f(\boldsymbol{\phi}^{(r)})}{\partial \boldsymbol{\phi}^{(r)}} \cdot f(\boldsymbol{\phi}^{(r)}),\\
h'(\boldsymbol{\phi}^{(r)})=\frac{\partial \mathrm{log} h(\boldsymbol{\phi}^{(r)})}{\partial \boldsymbol{\phi}^{(r)}} \cdot h(\boldsymbol{\phi}^{(r)}).
\end{align*}
We derive
\begin{equation}\label{derivative of w wrt phi}
\begin{split}
     \frac{\partial w_i^{(r)}}{\partial \boldsymbol{\phi^{(r)}}}&=
     \frac{\partial}{\partial \boldsymbol{\phi^{(r)}}}\frac{\pi_i^{(r)}f(\boldsymbol{\phi}^{(r)})}{\pi_i^{(r)}f(\boldsymbol{\phi}^{(r)})+(1-\pi_i^{(r)})h(\boldsymbol{\phi}^{(r)})}\\
     &=\frac{\pi_i^{(r)}(1-\pi_i^{(r)})\bigl[f'(\boldsymbol{\phi}^{(r)})h(\boldsymbol{\phi}^{(r)})-f(\boldsymbol{\phi}^{(r)})h'(\boldsymbol{\phi}^{(r)})\bigr]}{\bigl[\pi_i^{(r)}f(\boldsymbol{\phi}^{(r)})+(1-\pi_i^{(r)})h(\boldsymbol{\phi}^{(r)})\bigr]^2}\\
     &=\frac{\pi_i^{(r)}(1-\pi_i^{(r)})f(\boldsymbol{\phi}^{(r)})h(\boldsymbol{\phi}^{(r)})\bigl[ \partial\mathrm{log} f(\boldsymbol{\phi}^{(r)})/\partial\boldsymbol{\phi}^{(r)}-\partial\mathrm{log} h(\boldsymbol{\phi}^{(r)})/\partial\boldsymbol{\phi}^{(r)}\bigr]}{\bigl[\pi_i^{(r)}f(\boldsymbol{\phi}^{(r)})+(1-\pi_i^{(r)})h(\boldsymbol{\phi}^{(r)})\bigr]^2}
\end{split}
\end{equation}
where $\pi_i^{(r)}$ is calculated at $\alpha^{(r)}$ using formula \eqref{logit},  $f(\phi^{(r)})$ and $h(\phi^{(r)})$ are calculated at $\phi^{(r)}$ using formula \eqref{rewritten likelihood},  $\partial\mathrm{log}f(\phi^{(r)})/\partial\phi^{(r)}$ and $\partial\mathrm{log}h(\phi^{(r)})/\partial \phi^{(r)}$ at $\phi^{(r)}$ using formula \eqref{standard cox score for beta} and \eqref{standard cox score for hazard}.

With \eqref{derivative of w wrt phi}, \eqref{standard logit score},\eqref{first derivative b} and \eqref{first derivative hazard}, we can then obtain $\frac{\partial^2 Q(\boldsymbol{\theta}|\boldsymbol{\theta}^{(r)})}{\partial\alpha\partial\phi^{(r)T}}$, $\frac{\partial^2 Q(\boldsymbol{\theta}|\boldsymbol{\theta}^{(r)})}{\partial b_{kl} \partial\phi^{(r)T}}$ and $\frac{\partial^2 Q(\boldsymbol{b}|\boldsymbol{b}^{(r)})}{\partial\lambda_{0,j,kl}\partial\phi^{(r)T}}$, via the following relationships, which are similar as in \eqref{second term alpha alpha r}, \eqref{second term b alpha r} and \eqref{second term lambda alpha r}. The lengthy expressions are omitted here.

\begin{equation}\label{second term alpha phi r}
    -\frac{\partial^2 Q(\boldsymbol{\theta}|\boldsymbol{\theta}^{(r)})}{\partial\alpha\partial\phi^{(r)T}} = -\sum_{i=1}^{n}\frac{\partial w_i^{(r)}}{\partial \phi^{(r)}}Z_{i,\alpha},
\end{equation}
\begin{equation}\label{second term b phi r}
-\frac{\partial^2 Q(\boldsymbol{\theta}|\boldsymbol{\theta}^{(r)})}{\partial b_{kl}\partial\phi^{(r)T}} =  -\sum_{i=1}^{n} \sum_{G_i=0}^{1}\sum_{j:t_{(j)}\leq X_i}  (2G_{i}-1)\frac{\partial w_i^{(r)}}{\partial \phi^{(r)}} \bigl\{Z_{i,b}dN_{i,kl}(t_{(j)})-Y_{i,k}(t_{(j)})\lambda_{0,j,kl}Z_{i,b}^T\exp(Z_{i,b}^T b_{kl})\bigr\} ,
\end{equation}

\begin{equation}\label{second term lambda phi r}
 -\frac{\partial^2 Q(\boldsymbol{\theta}|\boldsymbol{\theta}^{(r)})}{\partial\lambda_{0,j,kl}\partial\phi^{(r)T}} =  - \sum_{i=1}^{n} \sum_{G_i=0}^{1}(2G_{i}-1)\frac{\partial w_i^{(r)}}{\partial \phi^{(r)}}\bigl\{\frac{dN_{i,kl}(t_{(j)})}{\lambda_{0,j,kl}}-Y_{i,k}(t_{(j)})\exp(Z_{i,b}^T b_{kl})\bigr\}. 
\end{equation}

Now that we have solved the second term $\frac{\partial^2 Q(\boldsymbol{\theta}|\boldsymbol{\theta}^{(r)})}{\partial\boldsymbol{\theta}\partial\boldsymbol{\theta}^{(r)T}}$ in \eqref{observed info formula}, we can sum it together with the first term derived above, $\frac{\partial^2 Q(\boldsymbol{\theta}|\boldsymbol{\theta}^{(r)})}{\partial\boldsymbol{\theta}\partial\boldsymbol{\theta}^T}$, to obtain the observed information matrix of the maximum likelihood estimator, through which we can have the standard errors of the parameter estimates from the EM algorithm.

\end{document}